\documentclass[10pt,conference]{IEEEtran}
\IEEEoverridecommandlockouts
\usepackage{cite}
\usepackage{microtype}
\usepackage{graphicx}
\usepackage{subcaption}
\usepackage{booktabs}
\usepackage{amsmath,amssymb,mathtools}
\usepackage{amsthm}
\usepackage{comment}
\usepackage{bm}
\usepackage{multirow}
\usepackage{hyperref}
\usepackage[capitalize,noabbrev]{cleveref}
\usepackage{algorithm}
\usepackage{algorithmic}
\usepackage{textcomp}
\usepackage{xcolor}
\def\BibTeX{{\rm B\kern-.05em{\sc i\kern-.025em b}\kern-.08em
    T\kern-.1667em\lower.7ex\hbox{E}\kern-.125emX}}
\newcommand{\todo}[1]{}
\renewcommand{\todo}[1]{{\color{red} TODO: {#1}}}

\theoremstyle{plain}

\theoremstyle{definition}

\theoremstyle{remark}

\begin{document}

\title{Toward Uncertainty-Aware and Generalizable Neural Decoding for Quantum LDPC Codes}

\author{
Xiangjun Mi, Frank Mueller \\
North Carolina State University, Raleigh, NC 27695, USA \\
\{xmi, fmueller\}@ncsu.edu
}

\maketitle

\begin{abstract}
Quantum error correction (QEC) is essential for scalable quantum
computing, yet decoding errors via conventional algorithms result in
limited accuracy (i.e., suppression of logical errors) and high
overheads, both of which can be alleviated by inference-based
decoders. To date, such machine-learning (ML) decoders lack two key
properties crucial for practical fault tolerance: reliable
uncertainty quantification and robust generalization to previously
unseen QEC codes. To address this gap, we propose a Quantum Bayesian
graph Attention decoder \textbf{(QuBA)} that enables expressive
error-pattern recognition alongside calibrated uncertainty
estimates. Building on QuBA, we further develop a multi-phase
training framework with enhanced cross-domain robustness enabling
decoding beyond the training set called Sequential Aggregate
Generalization under Uncertainty \textbf{(SAGU)}.  Experiments on
bivariate bicycle (BB) codes and their coprime variants demonstrate
that
(i) QuBA consistently outperforms the classical baseline belief propagation (BP), achieving up to a \emph{two orders of magnitude} reduction in logical error rate (LER) under confident-decision bounds on the coprime BB code $[[154,6,16]]$;
(ii) SAGU achieves decoding performance comparable to or even
outperforming QuBA's domain-specific training approach.
\end{abstract}

\section{Introduction}
Quantum error correction (QEC)~\cite {calderbank1996good} is an
essential paradigm that enables quantum computation at negligible
error rates over logical
qubits~\cite{kielpinski2002architecture,ye2023towards,he2025performance}. By
encoding a logical qubit into multiple physical qubits and measuring
parity-check syndromes, QEC can diagnose and correct logical errors
(bit and phase flips) without destroying a complex quantum
state~\cite{calderbank1996good,knill1997theory}. Even though physical
qubits are subject to noise (more frequent errors, currently around
$10^{-3}$), clever encoding into logical qubits reduces this error
rate algorithmically via error correction to where it becomes
negligible (up to $10^{-13}$) at the logical level. Quantum
low-density parity-check (LDPC) codes are a broad class of codes aimed
at high error-correction performance combined with high encoding
efficiency~\cite{gottesman1997stabilizer,tillich2013quantum}.  Early
quantum LDPC constructions included hypergraph product codes and
hyperbolic codes, which demonstrated that constant-rate quantum codes
with growing distance are
possible~\cite{freedman2002z2,zemor2009cayley,zeng2019higher,breuckmann2016constructions}. More
recent developments, such as balanced product codes and quantum Tanner
codes, have achieved even stronger asymptotic guarantees, with some
families proving to be good quantum codes, offering constant encoding
rates and linear distance
scaling~\cite{breuckmann2021balanced,leverrier2022quantum}. These
advances make quantum LDPC codes highly attractive for fault-tolerant
quantum computing, as they can dramatically reduce the physical qubit
overhead compared to surface codes while maintaining competitive
thresholds.  Among these, bivariate bicycle (BB)
codes~\cite{bravyi2024high}, published in \emph{Nature}, have
attracted particular attention for their balance between practicality
and asymptotic performance, which generalize classical bicycle codes
into two dimensions and achieve a threshold close to 0.8\%, comparable
to the surface code, but with substantially higher encoding
rates~\cite{postema2025existence}.

Efficient decoding is critical for realizing the benefits of quantum
codes with near-term quantum device technology. In decoding via
general Tanner graphs, iterative belief propagation (BP) decoding is
widely used due to its moderate computational complexity and high
degree of
parallelism~\cite{kschischang2002factor,poulin2008iterative}. However,
the abundance of \emph{short cycles} in Tanner graphs of quantum codes
can severely degrade BP performance~\cite{kovalev2013quantum}, while
\emph{degeneracy} (i.e., multiple distinct errors corresponding to the
same syndrome) can trap BP in symmetric belief
states~\cite{poulin2006optimal,panteleev2021degenerate}. To address
these limitations, several variants have been proposed.  Memory-based
BP introduces additional memory effects~\cite{kuo2022exploiting},
while SymBreak explicitly mitigates
degeneracy~\cite{yin2024symbreak}. Other improvements include guided
decimation~\cite{yao2024belief}, sliding window
decoding~\cite{gong2024toward}, automorphism-ensemble
decoding~\cite{koutsioumpas2025automorphism},
relay-BP~\cite{muller2025improved}, and speculative
approaches~\cite{wang2026fully}. Post-processing techniques have also
been developed, such as ordered statistics decoding
(OSD)~\cite{roffe2020decoding}, which improves performance but at
cubic computational cost and limited parallelism, as well as the more
recent localized statistics decoding
(LSD)~\cite{hillmann2025localized}, which offers a parallelizable
alternative. More recently, the QEC community has turned toward
advanced machine learning (ML)-based decoders, ranging from (recurrent
or graph) neural
networks~\cite{nachmani2018deep,liu2019neural,miao2023neural,baireuther2018machine}
and generative decoding~\cite{cao2025generative} to transformer
architectures~\cite{wang2023transformer,choukroun2024deep}, providing
a new perspective on decoding techniques.

Another crucial challenge in decoding is \emph{uncertainty}.  In QEC,
both the physical error process and the decoding model introduce
uncertainty.  \emph{Physical uncertainty} arises from the stochastic
nature of the Pauli noise channel, which produces different physical
error patterns even under identical circuit operations.  \emph{Model
  uncertainty} stems from limited training data, code degeneracy, and
imperfect generalization to unseen syndromes. Accurately representing
both types of uncertainty is essential. It allows the decoder to
output well-calibrated confidence estimates, improves robustness to
distribution shifts (e.g., changes in error rate), and supports
downstream decision-making such as hybrid decoding with ordered
statistics decoding (OSD)~\cite{roffe2020decoding}, etc.

Despite these advances, existing ML decoders
still face two critical limitations for practical fault
tolerance. First, they generally lack reliable uncertainty
  quantification, making it
difficult to assess confidence in decoding decisions or to design
adaptive hybrid strategies. Second, their generalization ability
across different quantum codes remains weak, as most approaches are
trained domain-specifically and fail to transfer to unseen codes or
varying noise conditions \emph{without any retraining}.

Motivated by these challenges, our contributions are summarized as follows:
\begin{itemize}
\item To the best of our knowledge, this work is the first in quantum
  ML-based decoding to jointly model and quantify both data and model
  uncertainty. To this end, we propose QuBA, which leverages Bayesian
  neural networks (BNNs) to represent predictive uncertainty, using
  Monte Carlo dropout at inference time to produce calibrated
  confidence estimates that enable adaptive decision-making. To better
  capture correlations in error syndromes, QuBA further integrates an
  attention mechanism into a graph neural network architecture,
  enhancing relational reasoning on Tanner graphs (see
  Sec. ~\ref{app:quba} ).
\item Beyond QuBA, and inspired by the DART
  paradigm~\cite{jain2023dart}, we design SAGU (Sequential Aggregate
  Generalization under Uncertainty), a cross-domain training framework
  that consists of three phases (see Sec.~\ref{SAGU}) designed to
  strengthen generalization across heterogeneous quantum codes, by
  exploiting the complementary strengths of diverse code architectures
  and training data distributions in the quantum decoding setting.
\item Together, QuBA and SAGU deliver not only improved decoding
  accuracy (i.e., suppression of logical errors) but also uncertainty
  awareness and strong cross-domain robustness. Experimental results
  across both standard and coprime BB codes illustrate the superiority
  of the proposed methods.  E.g., QuBA consistently
  outperforms the classical baseline BP, achieving, on average, \emph{one
    order of magnitude} reduction in LER and up to \emph{two orders of
    magnitude} under confident-decision bounds on the coprime BB code
  $[[154,6,16]]$. In addition, QuBA surpasses state-of-the-art neural
  decoders, providing roughly a \emph{one order of magnitude}
  advantage (e.g., for the larger BB code $[[756,16,\leq 34]]$) even
  under conservative (safe) decision bounds.
\end{itemize}

\section{Related Work}
\label{app:related_work}
Recent research on ML-based decoders has pushed the
boundaries of classical and quantum decoding. Broadly, neural decoders
can be grouped into two categories: model-based and model-free
approaches.

\subsubsection{Model-based decoding} 
Model-based approaches explicitly incorporate the Tanner graph
structure of quantum LDPC codes into the neural architecture. Two main
directions have emerged. \emph{The first} integrates belief
propagation (BP) with neural design by unfolding iterative BP updates
into a differentiable architecture, allowing the update rules to be
optimized through data-driven
training~\cite{nachmani2016learning,nachmani2018deep,nachmani2021autoregressive}.
This line of work was extended to QEC with neural BP decoders, which
adapt message-passing rules to handle degeneracy in quantum LDPC
codes~\cite{liu2019neural}. \emph{A second direction} leverages
message-passing mechanisms in GNNs, directly embedding Tanner graph
connectivity into learned aggregation and update functions. Recent
works demonstrated the effectiveness of GNN-based decoders for both
classical and quantum LDPC codes, with notable progress on quantum
LDPC decoding at
scale~\cite{gong2024graph,ninkovic2024decoding,maan2025machine}.
\cite{bhave2025hypernq}
introduces a hypergraph neural decoder with a two-stage
message-passing mechanism for hypergraph product codes, achieving up
to an 84\% reduction in LER compared to BP.  By combining structural
priors with trainable neural layers, model-based decoders achieve high
performance while usually retaining scalability and interpretability.

\subsubsection{Model-free decoding} 
In contrast, model-free methods treat decoding as a purely data-driven
task, without embedding explicit BP or Tanner graph mechanics. Early
works employed neural networks trained directly on error-syndrome
pairs to map syndromes to corrections for topological color codes,
demonstrating the feasibility of purely supervised decoders under
circuit-level noise~\cite{baireuther2019neural}. Subsequent advances
introduced attention-based architectures such as self-attention and
transformers, which are capable of capturing global correlations in
syndrome data and have shown strong decoding performance across
classical and quantum
codes~\cite{raviv2020perm2vec,choukroun2022error,cohen2025hybrid}.
Building on this line of work, a recurrent transformer decoder was
recently applied to BB codes, where a multi-stage training protocol
enabled effective decoding under circuit-level
noise~\cite{blue2026machine}. In addition, ~\cite{liu2025decoding}
proposes a diffusion-based decoder that directly predicts the LER from
detectors in STIM~\cite{gidney2021stim}, as in the transformer method,
and demonstrates performance improvements.  In parallel, GNNs have
also been explored in fully data-driven settings, where decoding is
formulated as a graph-classification problem, and the network directly
predicts the most likely logical error class ~\cite{lange2025data}.

\subsubsection{Our approach} 
Our proposed method, QuBA, belongs to the model-based category, as it
builds on neural message passing while augmenting it with Bayesian
attention mechanisms. 
Unlike prior model-based decoders, QuBA provides
explicit predictive uncertainty estimates and integrates attention to
capture heterogeneous syndrome-qubit interactions. Together with the
sequential training strategy SAGU, our framework achieves both
stronger error suppression and broader cross-domain generalization
than classical and related ML-based approaches.

\section{Background}
\begin{figure*}[htbp]
    \centering
    \includegraphics[width=\linewidth]{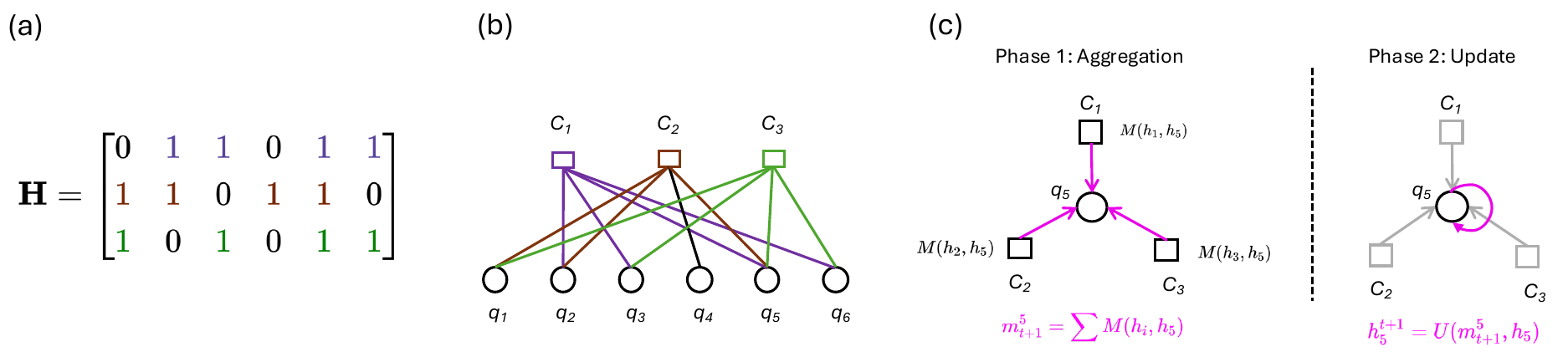}
    \caption{Relationship between belief propagation (BP) and message
    passing in graph neural networks (GNNs).
    \textbf{(a)}. A parity-check matrix.
    \textbf{(b)}. The Tanner graph constructed from the parity-check
    matrix, showing the connections between check nodes and variable
    nodes (physical qubits).
    BP operates by exchanging information between check and variable
    nodes (from check nodes to variable nodes, and vice versa) over
    $T$ iterations.
    \textbf{(c)}. A message-passing neural network (MPNN) on the
    Tanner graph. At each iteration, message passing proceeds in two
    steps. In the aggregation step, every node $v$ computes messages
    for its neighbors $u \in \mathcal{N}(v)$ by applying a learnable
    message function $M(\cdot)$. All incoming messages are aggregated
    at the receiving node using a permutation-invariant operator such
    as element-wise summation. In the update step, the hidden state of
    node $v$ is updated by an update function $U(\cdot)$ that combines
    the previous state with the aggregated messages. After $T$ such
    iterations, the hidden representation of each node reflects
    information from its $T$-hop neighborhood.}
    \label{fig:bp_vs_mpnn}
\vspace*{-\baselineskip}
  \end{figure*}

\subsection{Quantum Decoding}
In the stabilizer formalism~\cite{gottesman1997stabilizer}, a
$[[n,k,d]]$ quantum code, encoding $k$ logical qubits into $n$
physical qubits with a code distance $d$ (i.e., the minimum number of
physical qubit errors that can cause an undetectable logical error),
is defined by a stabilizer group
$\mathcal{S} = \langle S_1,\dots,S_{n-k} \rangle$ of commuting Pauli
operators.  The code space is the joint $+1$ eigenspace of all $S_i$.
An error $E$ anticommutes with a subset of stabilizers, producing a
binary syndrome vector
\begin{equation}
S_i = \begin{cases}
0, & ES_i = S_iE, \\
1, & ES_i = -S_iE,
\end{cases}
\quad j = 1,\dots,n-k.
\label{eq:quantum-syndrome}
\end{equation}
The quantum decoding problem is to find a correction
$E_{\mathrm{corr}}$ such that
\begin{equation}
E_{\mathrm{corr}} E \in \mathcal{S},
\label{eq:quantum-degeneracy}
\end{equation}
i.e., $E_{\mathrm{corr}}$ differs from $E$ by a stabilizer and thus
restores the code state up to a global phase. Given a syndrome
$\mathbf{s}$, a maximum-likelihood decoder selects
\begin{equation}
\hat{E} = \arg\max_{E \in \mathcal{P}_n} P(E \,|\, \mathbf{s}),
\label{eq:ml-quantum}
\end{equation}
where $\mathcal{P}_n$ is the $n$-qubit Pauli group. 
Graph-based
quantum decoders realize Eq.~\ref{eq:ml-quantum} as belief propagation
(BP) on the Tanner graph, where variable and check nodes exchange
belief information.

\subsection{Graph Neural Networks}

GNNs~\cite{scarselli2008graph,wu2020comprehensive,liu2022graph} extend
deep learning to graph-structured data by iteratively exchanging and
aggregating information between neighboring nodes. In
\emph{message-passing neural network} (MPNN)~\cite{gilmer2017neural},
the hidden state of each node $v$ at iteration $t$ is updated as
\begin{equation}
\begin{aligned}
\mathbf{m}_{v}^{(t)} &=
\square_{u \in \mathcal{N}(v)}
\psi^{(t)}\!\left(
  \mathbf{h}_v^{(t)}, \mathbf{h}_u^{(t)}, \mathbf{e}_{uv}
\right), \\
\mathbf{h}_v^{(t+1)} &=
\phi^{(t)}\!\left(
  \mathbf{h}_v^{(t)}, \mathbf{m}_v^{(t)}
\right),
\end{aligned}
\end{equation}
where $\psi^{(t)}$ is the message function, $\phi^{(t)}$ is the node
update function, $\mathbf{e}_{uv}$ encodes edge features, and
$\square$ denotes a permutation-invariant aggregation (e.g., sum,
mean, or max). This iterative scheme allows information to propagate
over multi-hop neighborhoods, enabling the network to capture both
local and global structural patterns.

Fig.~\ref{fig:bp_vs_mpnn} illustrates the relationship between BP and
neural message-passing networks on the Tanner graph.  For decoding,
the Tanner graph of a quantum LDPC code naturally provides the input
graph, where variable nodes correspond to physical qubits, check nodes
correspond to syndrome bits, and edges encode qubit-stabilizer
incidence. The GNN learns to propagate and transform messages in a
manner that approximates maximum-likelihood decoding, thereby
mitigating key limitations of BP in graphs with cycles. Specifically,
GNN-based message passing relies on learned state updates, whereas BP
performs iterative belief exchanges that accumulate toward a final
hard decision. In the presence of short cycles, these fixed update
rules in BP can give rise to trapping sets and degeneracy-induced
symmetries, leading to oscillatory behavior or biased belief
accumulation~\cite{raveendran2021trapping,chytas2024enhanced}.

\section{QuBA: A Quantum Bayesian Attention Decoder}\label{app:quba}
\begin{figure*}[!htbp]
    \centering
    \includegraphics[width=\linewidth]{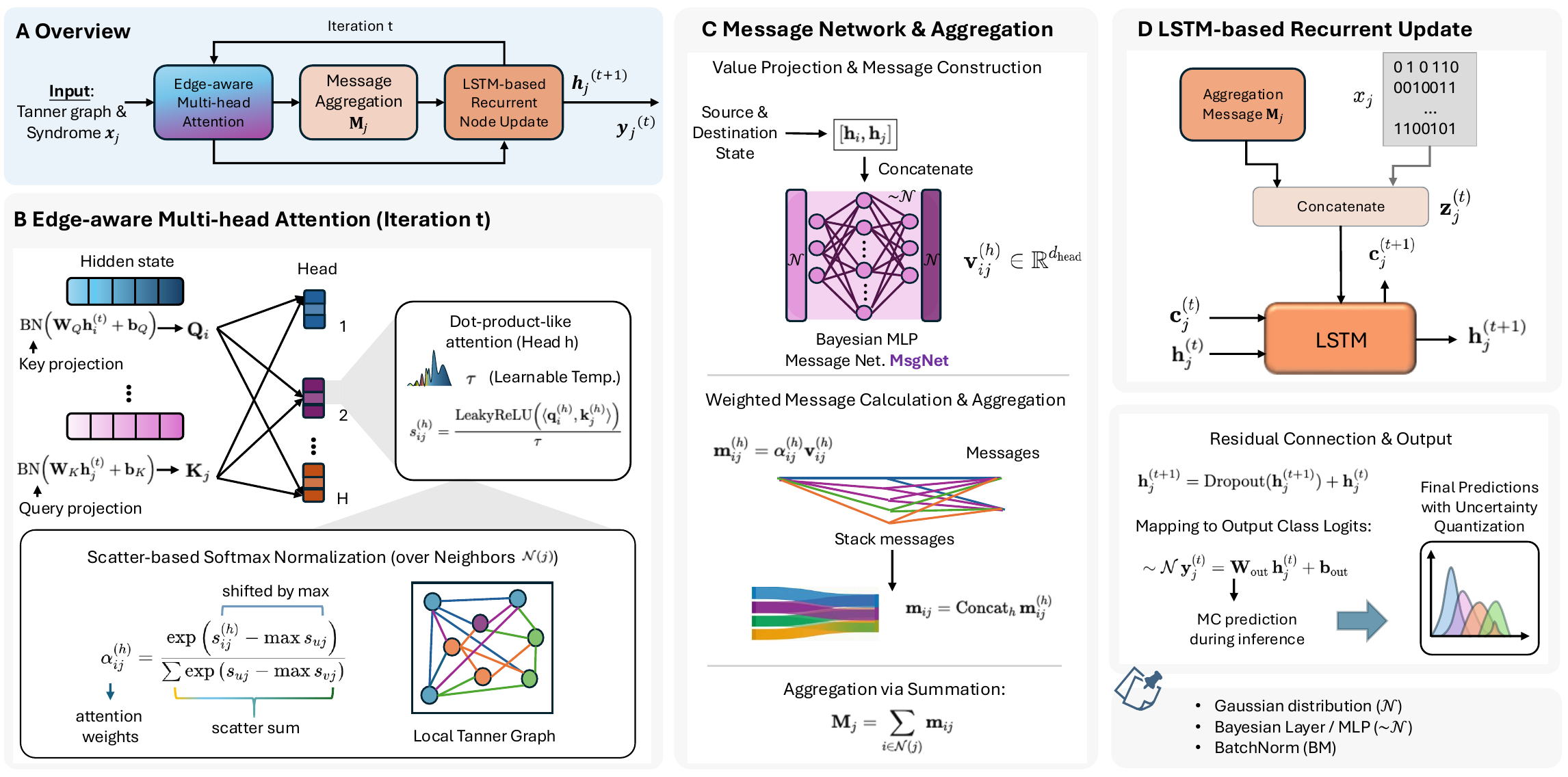}
    \caption{Architecture of QuBA.}
    \label{fig:quba}
\vspace*{-\baselineskip}
\end{figure*}

\subsection{Uncertainty Representation}

To enable uncertainty-aware decoding, we adopt a Bayesian neural
network (BNN) formulation for our GNN-based decoder. Unlike
conventional deep neural networks (DNNs) that learn a single point
estimate for each parameter, in a BNN every model parameter
$\theta \in \Theta$ is treated as a random variable rather than a
fixed value. A prior distribution $p(\theta)$, chosen as a standard
Gaussian $\mathcal{N}(0,1)$, encodes our initial belief about these
parameters before observing any data. Given a training dataset
$ \mathcal{D} = \{ (\mathbf{s}_i, \mathbf{e}_i) \}_{i=1}^N$, where
$\mathbf{s}_i$ denotes the measured stabilizer syndrome and
$\mathbf{e}_i$ is the corresponding physical error pattern on data
qubits, the posterior distribution over parameters is obtained via
Bayes’ theorem:
\begin{equation}
p(\theta \mid \mathcal{D}) 
= \frac{p(\mathcal{D} \mid \theta) \, p(\theta)}{p(\mathcal{D})} 
= \frac{\prod_{i=1}^N p(\mathbf{e}_i \mid \mathbf{s}_i, \theta) \, p(\theta)}{p(\mathcal{D})}.
\label{eq:bayes_posterior_qec}
\end{equation}

For a new measured syndrome $\mathbf{s}^*$, the predictive
distribution over the corresponding physical error pattern
$\mathbf{e}^*$ is obtained by marginalizing over the posterior as
\begin{equation}
p(\mathbf{e}^* \mid \mathbf{s}^*, \mathcal{D}) 
= \int_{\theta} p(\mathbf{e}^* \mid \mathbf{s}^*, \theta) \; p(\theta \mid \mathcal{D}) \; d\theta,
\label{eq:predictive_qec}
\end{equation}
and the prediction of such a marginal distribution incorporates both
data and model uncertainties~\cite{abdar2021review}.

In practice, however, the exact posterior $p(\theta \mid \mathcal{D})$
in Eq.~\ref{eq:bayes_posterior_qec} is intractable for gradient-based
optimization, since it requires integration over the entire parameter
space $\Theta$. To make learning feasible, we approximate the
posterior with a factorized Gaussian variational distribution
$q_\phi(\theta)$~\cite{graves2011practical,blundell2015weight,louizos2016structured},
parameterized by variational parameters $\phi$. This approximation is
optimized by minimizing the Kullback–Leibler (KL) divergence
\begin{equation}
\min_{\phi} \; \mathrm{KL}\!\left(q_\phi(\theta) \,\|\, p(\theta \mid \mathcal{D})\right) 
= \min_{\phi} \int_{\theta \in \Theta} q_\phi(\theta) \log \frac{q_\phi(\theta)}{p(\theta \mid \mathcal{D})} \, d\theta.
\end{equation}

Under the Gaussian assumption in each BNN layer, 
this KL term has the closed-form expression
\begin{equation}
KL = \frac{1}{2} \sum_j \left[ \frac{\sigma_j^2}{\sigma_p^2} + \frac{\mu_j^2}{\sigma_p^2} - 1 - \log\frac{\sigma_j^2}{\sigma_p^2} \right],
\label{eq:kl_closedform_qec}
\end{equation}
where $(\mu_j, \sigma_j^2)$ are the variational parameters of
$q_\phi(\theta)$ and $\sigma_p^2$ is the prior variance.  The total KL
regularizer is summed over all Bayesian layers in the decoder,
ensuring posterior distributions remain close to the prior.

\paragraph{Monte Carlo (MC) prediction}
At inference, we perform $M$ independent MC forward passes.
During each pass $m$, the decoder evaluates the same syndrome
$\mathbf{s}$ while drawing fresh layerwise weight samples from
$q_\phi(\theta)$ at each internal iteration (i.e., weights are
resampled within the forward). This produces predictions
$\hat{\mathbf{e}}^{(m)}$, whose empirical mean and variance are
\begin{equation}
\hat{\mu} = \frac{1}{M}\sum_{m=1}^M \hat{\mathbf{e}}^{(m)}, \qquad
\hat{\sigma}^2 = \frac{1}{M}\sum_{m=1}^M \bigl(\hat{\mathbf{e}}^{(m)}-\hat{\mu}\bigr)^2 .
\label{eq:mc_estimates_qec}
\end{equation}
A $95\%$ confidence interval for each predicted error probability is
then approximated by
$ CI^{0.95} \approx \hat{\mu} \pm 2 \hat{\sigma}.$
This MC-based prediction procedure captures epistemic uncertainty
through weight sampling and predictive variability from the decoder's
output distribution, both of which are crucial for robust and reliable
QEC decoding.

\subsection{Decoder Design}
Our decoder is a graph neural network that integrates 
(i) \emph{edge-aware multi-head attention} and 
(ii) \emph{LSTM-based recurrent state updates} 
with Bayesian parameterization for uncertainty quantification.

\paragraph{Node initialization}
Each node $i$ begins from a shared learnable embedding
\begin{equation}
\mathbf{h}_i^{(0)} = \mathbf{e}_0 \in \mathbb{R}^{d_h},
\end{equation}
where $d_h$ is the hidden dimension. This provides a uniform
initialization for iterative message passing.

\paragraph{Edge-aware multi-head attention}
At iteration $t$, hidden states are projected into queries and keys
using Bayesian linear layers with BatchNorm (BN)
\begin{equation}
\mathbf{Q}_i = \mathrm{BN}\!\left(\mathbf{W}_Q \mathbf{h}_i^{(t)} + \mathbf{b}_Q\right), \quad
\mathbf{K}_j = \mathrm{BN}\!\left(\mathbf{W}_K \mathbf{h}_j^{(t)} + \mathbf{b}_K\right),
\end{equation}
where $\mathbf{W}_Q,\mathbf{W}_K \in \mathbb{R}^{d_h \times (H d_{\text{head}})}$. 
The vectors are reshaped into $H$ heads of dimension $d_{\text{head}}$.

For each edge $(i \!\to\! j)$ and head $h$, the scaled dot-product
attention score is
\begin{equation}
s_{ij}^{(h)} = \frac{\mathrm{LeakyReLU}\!\left(\langle \mathbf{q}_i^{(h)}, \mathbf{k}_j^{(h)}\rangle\right)}{\tau},
\end{equation}
where $\tau$ is a learnable temperature.
To stabilize training, scores are shifted by the maximum value at each
destination and normalized with a scatter-based softmax
\begin{equation}
\alpha_{ij}^{(h)} = 
\frac{\exp(s_{ij}^{(h)} - \max_{u\in \mathcal{N}(j)} s_{uj}^{(h)})}
{\sum_{u\in \mathcal{N}(j)} \exp(s_{uj}^{(h)} - \max_{v\in \mathcal{N}(j)} s_{vj}^{(h)})}.
\end{equation}

\paragraph{Message network}
Values are produced by a deep Bayesian MLP operating on concatenated
source and destination states
\begin{equation}
\mathbf{v}_{ij} = \mathrm{MsgNet}\!\left([\mathbf{h}_i^{(t)}, \mathbf{h}_j^{(t)}]\right) 
\in \mathbb{R}^{H d_{\text{head}}},
\end{equation}
with per-head values
$\mathbf{v}_{ij}^{(h)} \in \mathbb{R}^{d_{\text{head}}}$,
and messages are then scaled by attention weights
\begin{equation}
\mathbf{m}_{ij}^{(h)} = \alpha_{ij}^{(h)} \mathbf{v}_{ij}^{(h)}, \quad
\mathbf{m}_{ij} = \mathrm{Concat}_h \mathbf{m}_{ij}^{(h)}.
\end{equation}

\paragraph{Aggregation}
Finally, the messages are aggregated by summation over incoming edges
\begin{equation}
\mathbf{M}_j = \sum_{i \in \mathcal{N}(j)} \mathbf{m}_{ij}.
\end{equation}

\paragraph{LSTM-based recurrent update}
Each node update concatenates aggregated messages with static node
inputs $\mathbf{x}_j$
\begin{equation}
\mathbf{z}_j^{(t)} = [\mathbf{M}_j, \mathbf{x}_j].
\end{equation}
The Long Short-Term Memory (LSTM) cell then updates the hidden and
cell states
\begin{equation}
(\mathbf{h}_{j}^{(t+1)}, \mathbf{c}_{j}^{(t+1)}) 
= \mathrm{LSTM}\!\left(\mathbf{z}_j^{(t)}, (\mathbf{h}_j^{(t)}, \mathbf{c}_j^{(t)})\right),
\end{equation}
where $\mathbf{h}_j^{(t+1)}$ denotes the hidden state of node $j$ at
iteration $t\!+\!1$ (short-term representation), while
$\mathbf{c}_j^{(t+1)}$ denotes the corresponding cell state (long-term
representation), which preserves long-range dependencies across
multiple syndrome updates. 

A residual connection with dropout stabilizes the dynamics
\begin{equation}
\mathbf{h}_j^{(t+1)} = \mathrm{Dropout}(\mathbf{h}_j^{(t+1)}) + \mathbf{h}_j^{(t)}.
\end{equation}

\paragraph{Final output}
At each iteration, hidden states are mapped to class logits via a
Bayesian linear output layer
\begin{equation}
\mathbf{y}_j^{(t)} = \mathbf{W}_{\text{out}} \mathbf{h}_j^{(t)} + \mathbf{b}_{\text{out}}.
\end{equation}

Overall, this design enables the decoder to (i) adaptively weight
syndrome-qubit interactions through attention, (ii) propagate
parameter uncertainty through Bayesian layers, and (iii) maintain
long-range temporal consistency with recurrent memory in quantum
decoding.  Together, these mechanisms provide robustness to degeneracy
and improved generalization on large Tanner graphs with circular
dependencies. Fig.~\ref{fig:quba} shows an overview of our proposed
QuBA and detailed modules in it.

\subsection{Loss Function}
In QEC, let $\mathbf{e} \in \{0,1\}^{2n}$ denote the true Pauli error
in binary symplectic form, and let $\hat{\mathbf e}$ denote the
decoder's predicted correction. The residual error after applying the
correction is
$ \mathbf e_{\mathrm{tot}}=\mathbf e+\hat{\mathbf e}\pmod 2.  $
Decoding succeeds when this residual is trivial up to stabilizers, so
that it produces neither syndrome defects (i.e., stabilizer
violations) nor a nontrivial logical effect on the encoded state.

\paragraph{Logical-consistency loss}
To encourage globally valid corrections during training, QuBA
introduces a differentiable loss term based on the residual
error. Since exact mod-2 parity checks are non-differentiable, the
implementation replaces them with the smooth surrogate
$ f(x)=\left|\sin\left(\frac{\pi}{2}x\right)\right|
$~\cite{liu2019neural}.  The decoder outputs logits over the four
Pauli classes $\{I,X,Z,Y\}$ for each data qubit.  These logits are
first converted by a softmax into class probabilities.  The resulting
probabilities are then used to form soft X- and Z-components of the
predicted correction, which are combined with the ground-truth X- and
Z-components of $\mathbf e$ to obtain a soft version of the residual
$\mathbf e_{\mathrm{tot}}$. The resulting soft residual is projected
onto code-constraint subspaces using the orthogonal-complement bases
$H_X^\perp$ and $H_Z^\perp$, yielding the differentiable loss term
$\mathcal L_{\mathrm{LER}}$. This term penalizes residual errors that
are inconsistent with the code-level constraint structure.

\paragraph{Error cross-entropy loss}
We additionally supervise the predicted Pauli label on each data qubit
with a standard cross-entropy loss, denoted
$\mathcal{L}_{\mathrm{CE,e}}$. This term promotes accurate local
identification of the error type.

\paragraph{Syndrome cross-entropy loss}
Because the QuBA forward pass does not explicitly enforce syndrome
consistency, we also apply a cross-entropy loss
$\mathcal{L}_{\mathrm{CE,s}}$ on the syndrome-node outputs. This term
encourages agreement between the predicted and measured syndromes.

\paragraph{Overall objective}
The final objective averages the task losses across the $T$ decoding
iterations and adds the KL regularizer from the Bayesian neural
network formulation:
  \begin{equation}\label{loss}
  \begin{aligned}
  \mathcal{L}(\theta)
  &= \frac{1}{T}\sum_{t=1}^T
  \Big(
  \mathcal{L}^{(t)}_{\mathrm{LER}}
  + \tfrac{1}{2}\mathcal{L}^{(t)}_{\mathrm{CE,e}}
  + \tfrac{1}{2}\mathcal{L}^{(t)}_{\mathrm{CE,s}}
  \Big) \\
  &\quad
  + \beta(\tau)\,
  \mathrm{KL}\!\left(
  q_\phi(\theta)\,\|\,p(\theta)
  \right),
  \end{aligned}
  \end{equation}
  where $\beta(\tau)$ is annealed during training to gradually introduce
  Bayesian regularization.

In summary, this composite loss encourages the decoder to generate
corrections that are not only statistically accurate but also
logically valid under the stabilizer formalism, while explicitly
modeling epistemic uncertainty through Bayesian parameterization. 

\section{SAGU: Generalizable Training}\label{SAGU}

To address the challenge of learning a unified decoding strategy
across heterogeneous quantum code families, we develop a framework
that jointly captures variations in code constructions and data
distributions to ease effects such as trapping sets and quantum
degeneracy by leveraging the diversity across different codes. To this
end, we formalize this setting through a Bayesian graph decoding
framework that is inherently robust to domain shifts across quantum
codes, and introduce SAGU (\emph{Sequential Aggregate Generalization
  under Uncertainty}). SAGU enables principled, uncertainty-aware
generalization across diverse code families.

Within our SAGU framework, training is organized into three phases:
\emph{Warm-up}, \emph{Diversify-Aggregate}, and \emph{Consolidation}.
Each model is trained on distinct datasets in the corresponding phase,
while the training and validation sets are of equal size in every
phase, i.e.,
$ |\mathcal{D}_{\mathrm{warm}}| = |\mathcal{D}_{\mathrm{con.}}| =
\sum_{k=1}^M |\mathcal{D}_k|, $ where
\( \mathcal{D}_{\mathrm{warm}} \) and
\( \mathcal{D}_{\mathrm{con.}} \) denote the training or validation
sets used in the warm-up and consolidation phases, respectively, and
\( \mathcal{D}_{k} \) indicates the training or validation set for the
\(k\)-th model in the diversify-aggregate phase.  Note that the three
phases use the same loss objective defined in Eq.~\ref{loss}. We refer
to the trained models (i.e., the decoders for specific QEC codes) in
the three phases as the \emph{starting domain}, the \emph{diversity
  domain}, and the \emph{aggregation domain}. Collectively, these
three domains are referred to as \emph{in-domain}, while models
outside of them are considered \emph{out-of-domain}. More
specifically, the details and roles of each phase are described in the
following paragraphs, and the three-phase training procedure is
summarized in Alg.~\ref{alg:dp-dart}

\textbf{Warm-up:} We first optimize a single decoder $f_{\theta}$,
serving as the \emph{starting domain}, on
$\mathcal{D}_{\mathrm{warm}}$ for $E_m$ epochs using AdamW with a
phase-specific StepLR schedule.  This stage yields parameters
$\theta_{\mathrm{start}}$ that capture general decoding structure and
serve as the initialization and the starting point for all
domain-specific models. Typically, the starting point is a small QEC
code, i.e., one with a smaller code distance that can correct fewer
errors.
\textbf{Diversify-Aggregate:} We instantiate $M$ \emph{diversity
  domain} decoders $\{f_{\theta_k}\}_{k=1}^M$ with the trained model
in the previous phase $\theta_k\!\gets\!\theta_{\mathrm{start}}$.  For
each epoch $\tau\in[E_w, E_m)$, every model is trained independently
on its domain $\mathcal{D}_k$ with its own optimizer and StepLR.
After every $\lambda$ epochs (or at $\tau{=}\,E_m{-}1$, ), parameters
are synchronized by a weighted average
$\bar{\theta}=\sum_{k=1}^M w_k\,\theta_k$ with $\sum_k w_k{=}1$, where
the weights bias aggregation toward harder domains (larger $d_k$).
This balances domain-specific specialization with cross-domain sharing
of structural knowledge, including patterns arising from different
code structures and quantum degeneracy.
\textbf{Consolidation:} The final centralized $\bar{\theta}$
initializes a single \emph{aggregation domain} decoder.  We fine-tune
it for the remaining epochs $[E_m, E_{\mathrm{tot}})$ on the target
dataset (same size as warm-up) with a reduced learning rate and
StepLR, selecting checkpoints by validating logical-error metrics and
applying early stopping when the logical error rate (LER) fails to
improve within a patience window or when the LER completely converges
(i.e., reaches zero). Fig.~\ref{fig:sagu} presents the SAGU pipeline.

\begin{algorithm}[htbp]
\caption{Sequential Aggregate Generalization under Uncertainty (SAGU)}
\label{alg:dp-dart}
\small
\setlength{\itemsep}{0pt}
\setlength{\baselineskip}{0.95\baselineskip}
\begin{algorithmic}
\STATE \textbf{Input:}
Datasets $\mathcal{D}_{\mathrm{warm}}, \mathcal{D}_{\mathrm{con.}}, \{\mathcal{D}_k\}_{k=1}^M$;
aggregation weights $\{w_k\}_{k=1}^M$;
epoch budgets $E_w < E_m \le E_{\mathrm{tot}}$;
aggregation interval $\lambda$.
\STATE \textbf{Output:} Final decoder parameters $\theta_{\mathrm{final}}$.

\STATE \textbf{1. Warm-up Phase:}
\STATE Initialize starting-domain decoder $f_{\theta_{\mathrm{start}}}$.
\FOR{$\tau = 0$ \textbf{to} $E_w-1$}
  \STATE Train $f_{\theta_{\mathrm{start}}}$ for one epoch on $\mathcal{D}_{\mathrm{warm}}$.
\ENDFOR

\STATE \textbf{2. Diversify--Aggregate Phase:}
\STATE Initialize $M$ diversity domain decoders $\{f_{\theta_k}\}_{k=1}^M$:
$\theta_k \gets \theta_{\mathrm{start}}$, $k=1,\dots,M$.
\FOR{$\tau = E_w$ \textbf{to} $E_m-1$}
  \FOR{$k = 1$ \textbf{to} $M$}
    \STATE Train $f_{\theta_k}$ on domain $\mathcal{D}_k$ for one epoch.
  \ENDFOR
  \IF{$(\tau+1-E_w) \bmod \lambda = 0$ \textbf{or} $\tau = E_m-1$}
    \STATE Aggregate:
    $\bar{\theta} \gets \sum_{k=1}^M w_k \theta_k$.
    \STATE Synchronize: $\theta_k \gets \bar{\theta}$ for all $k$.
  \ENDIF
\ENDFOR

\STATE \textbf{3. Consolidation Phase:}
\STATE Initialize aggregation-domain decoder $f_{\theta_{\mathrm{final}}}$ with $\bar{\theta}$.
\FOR{$\tau = E_m$ \textbf{to} $E_{\mathrm{tot}}-1$}
  \STATE Fine-tune on $\mathcal{D}_{\mathrm{con.}}$ with reduced learning rate.
\ENDFOR

\STATE \textbf{Return} $\theta_{\mathrm{final}}$.
\end{algorithmic}
\end{algorithm}

\begin{figure*}[htbp]
    \centering
    \includegraphics[width=0.94\linewidth, height=7.4cm, keepaspectratio]{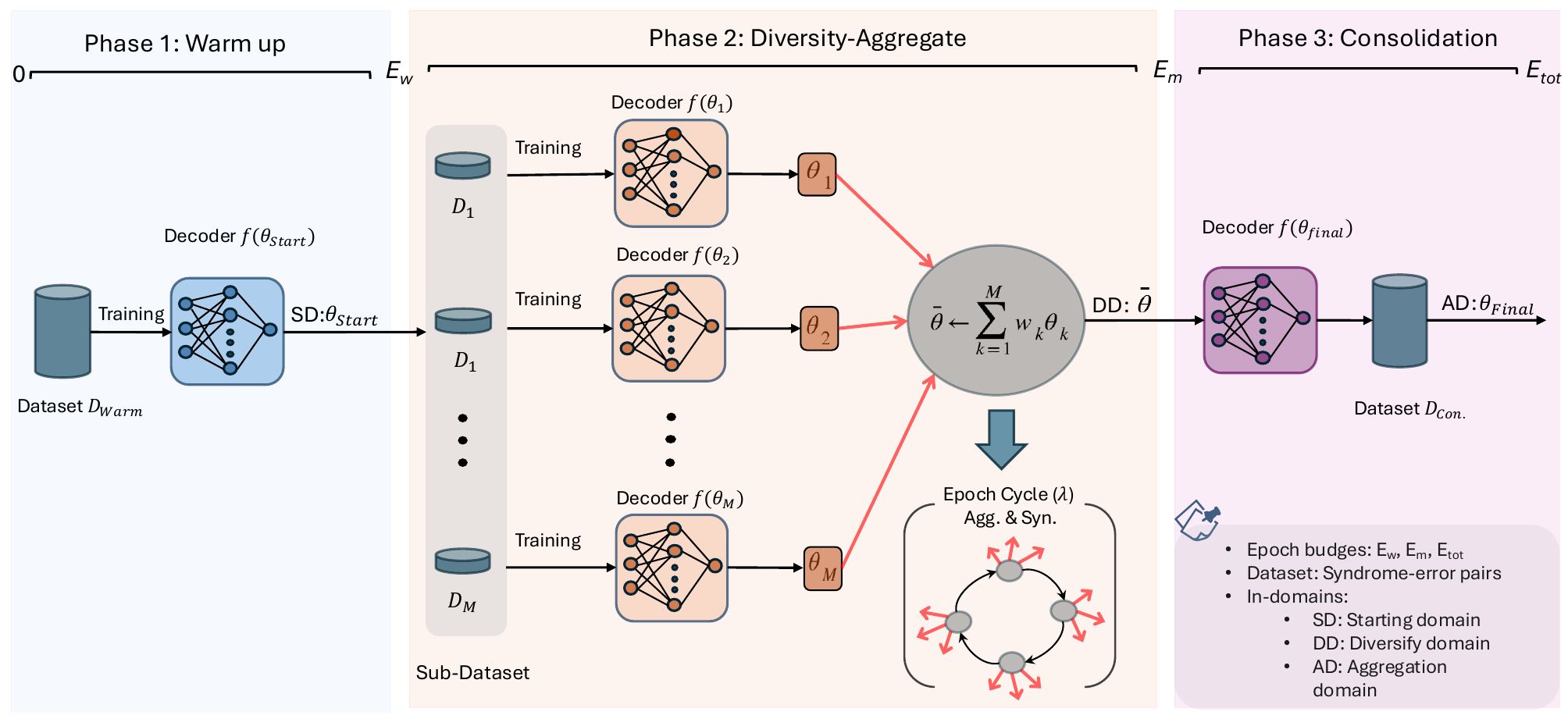}
    \caption{A diagram of the SAGU pipeline.}
    \label{fig:sagu}
\vspace*{-\baselineskip}
\end{figure*}
\section{Experiments}

  \subsection{Data}
  Each model, including those trained on BB codes, coprime BB codes, and
  within the SAGU framework, was trained using paired error-syndrome data
  generated under a capacity noise model. Specifically, we use an
  i.i.d.\ single-qubit depolarizing Pauli channel with noiseless syndrome
  extraction. For a physical error rate $p$, each data qubit remains
  error-free with probability $1-p$, and undergoes an $X$, $Y$, or $Z$
  error with probabilities $p/3$, $p/3$, and $p/3$, respectively.

  During training, $p$ is sampled uniformly from $[0,p_{\max}]$, where
  $p_{\max}$ is chosen close to the theoretical noise threshold of the
  corresponding QEC code. During testing, $p$ is fixed to the reported
  physical error rate. The syndrome/input node features are represented
  using a one-hot encoding, while the error labels are represented by four
  integer classes corresponding to $\{I,X,Z,Y\}$. Across all code families
  and models, results were obtained from over 1,000 syndrome samples.

  \subsection{Error Assumptions}
  \label{app:error_assump}
  In the literature, two types of error decoding methods are commonly
  considered, namely \emph{uncorrelated decoding} and \emph{correlated
  decoding}~\cite{maan2026decoding}. The former decodes the $X$ and $Z$
  error components separately, and therefore does not explicitly model
  same-qubit correlations arising from $Y$ errors, which may lead to
  suboptimal performance under depolarizing noise.

  QuBA employs a correlated decoding strategy by using a four-class error
  alphabet corresponding to $\{I,X,Z,Y\}$. In the implementation, the
  classes are encoded as $0,1,2,3$, where class $3$ denotes a $Y$ error.
  This class is interpreted as the simultaneous occurrence of its Pauli
  components, since $Y=iXZ$ up to a global phase. Thus, when evaluating
  residual and logical errors, class $3$ contributes to both the $X$ and
  $Z$ components.
  
  Although errors are independent across different qubits, the induced
  binary $X$ and $Z$ components on the same qubit are correlated under
  depolarizing noise. In particular, both $\Pr(e_X=1)$ and
  $\Pr(e_Z=1)$ are equal to $2p/3$, while
  $\Pr(e_X=e_Z=1)=p/3$. This same correlated noise model is used for both
  training and testing.

  \subsection{Baselines}
  We benchmark two classical decoding methods, BP~\cite{poulin2008iterative}
  and BP-OSD~\cite{roffe2020decoding}, as well as a state-of-the-art neural
  decoder, Astra~\cite{maan2025machine}, on both BB codes~\cite{bravyi2024high}
  and coprime BB codes~\cite{wang2024coprime,wang2026coprime}. All
  methods are evaluated under the same capacity depolarizing Pauli channel
  described above. We then compare these baselines with our proposed
  decoders, QuBA and SAGU, evaluating performance both with and without
  OSD post-processing.

\subsection{Training Details}
\paragraph{Model hyperparameters}  
We first report the set of hyperparameters that are shared across all
model variants. Model-specific hyperparameters are then detailed
separately. Tab.~\ref{tab:bb_coprime_hparams_full} summarizes the
training configurations for BB and coprime BB codes, while
Tab.~\ref{tab:training_phases_hparams} provides the hyperparameters
for training the SAGU model.
All models are trained using PyTorch's distributed data parallel (DDP)
framework on a workstation
equipped with three A5000 Ada GPUs. Each node is initialized with
\( n_{\text{node\_inputs}} = 4 \) input features, and the final
Bayesian output layer produces predictions of size
\( n_{\text{node\_outputs}} = 4 \).  The number of attention heads is
set to 4. Dropout rates are fixed at 0.1 for both the message network
(MsgNet) and the LSTM. The maximum physical error rate is
\( p_{\max} = 0.15 \), and the test error rate is fixed at
\( p_{\text{test}} = 0.05 \).  An AdamW optimizer is employed, with
weight decay \( 10^{-4} \) and learning rates specified separately in
the corresponding tables for each model.  The batch size is set to
16. The loss function, described in Eq.~\ref{loss}, incorporates KL
annealing over 10 epochs with a final scaling factor of \( 10^{-5}
\). Training is performed using automatic mixed precision (AMP), and
gradient clipping is applied with threshold
\( \lVert g \rVert \leq 1.0 \). Early stopping is triggered when the
logical error rate (LER) reaches zero, or if no improvement in LER is
observed over 20 consecutive epochs.

\begin{figure}[!htbp]
\centering
\renewcommand{\arraystretch}{0.9} 
\includegraphics[width=\linewidth, height=4.4cm, keepaspectratio]{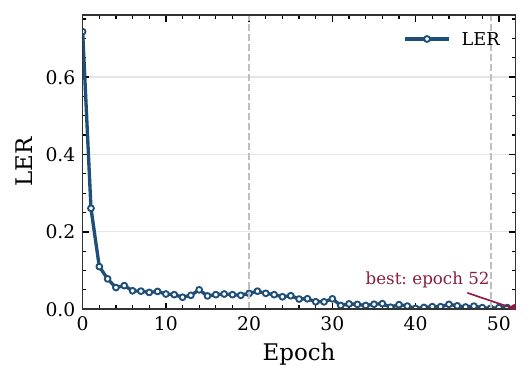}
\caption{LER across
training epochs for SAGU. For epochs 20--49, the reported value is averaged
over 3 decoders. Vertical dashed lines mark epochs 20 and 49.}
\label{fig:sagu_ler_train}
\end{figure}

\paragraph{SAGU training process}
Fig.~\ref{fig:sagu_ler_train} illustrates the training dynamics of
SAGU on the validation set. Epochs 0–19 correspond to the warm-up
phase, while epochs 50–52 shown in the figure fall within the
consolidation phase. The intermediate epochs correspond to the
diversity-aggregate phase. The results show that each phase
contributes to a steady reduction in the LER, culminating in full
convergence by epoch 52, at which point training is early
stopped. These observations validate the effectiveness and design
rationale of SAGU.

\begin{table*}[!htbp]
\centering
\begin{tabular}{c|c|c|c|c|c|c|c}
\hline
Code & $n_{\text{iters}}$ & $n_{\text{node}}$ & $n_{\text{edge}}$ & $\text{MsgNet size}$ & Training size & Validation size & LR \\
\hline
\multicolumn{8}{c}{BB codes} \\
\hline
$[[90, 8, 10]]$   & 40 & 64 & 32 & 256 & 50{,}000  & 3{,}000  & $5\times 10^{-4}$ \\
$[[144, 12, 12]]$ & 50 & 32 & 32 & 256 & 80{,}000  & 4{,}000  & $5\times 10^{-4}$ \\
$[[288, 12, 18]]$ & 65 & 64 & 32 & 128 & 100{,}000 & 5{,}000  & $5\times 10^{-4}$ \\
$[[756, 16, \leq34]]]$ & 50 & 64 & 32 & 128 & 50{,}000  & 3{,}000  & $5\times 10^{-4}$ \\
\hline
\multicolumn{8}{c}{Coprime BB codes} \\
\hline
$[[30, 4, 6]]$    & 40 & 32 & 32 & 256 & 30{,}000  & 1{,}000  & $5\times 10^{-4}$ \\
$[[154, 6, 16]]$  & 60 & 64 & 32 & 128 & 100{,}000 & 5{,}000  & $5\times 10^{-4}$ \\
\hline
\end{tabular}
\caption{Hyperparameters for BB and Coprime BB codes during training. }
\label{tab:bb_coprime_hparams_full}
\end{table*}

\begin{table*}[htbp]
\centering
\begin{tabular}{l|c|c|c|c|c|c|c|c}
\hline
Phase & BB code & $n_{\text{iters}}$ & $n_{\text{node}}$ & $n_{\text{edge}}$ & MsgNet size & Training size & Validation size & LR \\
\hline
Warm-up & $[[72,12,6]]$   & 35 & 64 & 32 & 128 & 24{,}000 & 1{,}200 & $5\!\times\!10^{-4}$ \\
\hline
\multirow{3}{*}{Diversify-Aggregate} 
  & $[[90,8,10]]$   & 40 & 64 & 32 & 128 & 6{,}000  & 300   & $5\!\times\!10^{-4}$ \\
  & $[[144,12,12]]$ & 50 & 64 & 32 & 128 & 8{,}000  & 400   & $5\!\times\!10^{-4}$ \\
  & $[[288,12,18]]$ & 65 & 64 & 32 & 128 & 10{,}000 & 500   & $5\!\times\!10^{-4}$ \\
\hline
Consolidation & $[[288,12,18]]$ & 50 & 64 & 32 & 128 & 24{,}000 & 1{,}200 & $1\!\times\!10^{-4}$ \\
\hline
\end{tabular}
\caption{Hyperparameters across the training phases in the SAGU
  schedule.  The total training budget is $E_{\text{total}}=90$
  epochs, with a warm-up $E_w=20$ and a mid-phase $E_m=50$.
  Aggregation occurs every $\lambda=10$ epochs using weighted
  averaging (weights $[0.1,0.2,0.7]$).  A StepLR scheduler is used
  with warm-up step $\lfloor 2/3E_w \rfloor$, domain step
  $\lfloor 2/3(E_m{-}E_w) \rfloor$, final step $=10$, and
  $\gamma=0.5$. }
\label{tab:training_phases_hparams}
\vspace*{-1.5\baselineskip}
\end{table*}

\subsection{Comparison Details}
\label{app:comparative_settings}

For a fair comparison, Astra and QuBA were trained using identical
hyperparameters and the same training and test datasets. Comparisons
with BP were performed on the same test sets. To balance computational
depth, we allowed twice as many message-passing iterations in the
learned models (Astra, QuBA, and SAGU) as in BP, since BP performs
bidirectional message updates, while the learned models employ
unidirectional message passing, i.e., from syndrome nodes to variable
nodes. Finally, in experiments with OSD post-processing (applied
independently to both $X$- and $Z$-decoders), all methods used the
same configuration given by \texttt{schedule} = \texttt{serial},
\texttt{bp\_method} = \texttt{ms}, \texttt{ms\_scaling\_factor} =
0.725, and \texttt{osd\_method} = \texttt{osd0}. Furthermore, we
assess SAGU under a domain-shift protocol with four BB domains
(starting, diversity and aggregation domains), and an out-of-domain
evaluation. We compare our approach against the baseline methods BP
and BP-OSD. To evaluate the performance gains of SAGU over the QuBA
code-specific training, each domain code is trained using the
hyperparameters listed in Tab.~\ref{tab:training_phases_hparams}, with
fixed training and testing dataset sizes of 24,000 and 1,200,
respectively, for all codes. For the out-of-domain BB code
\([[756,16,\leq34]]\), the hyperparameters are identical to those of
the consolidation phase, except that the learning rate (LR) is set to
\(5 \times 10^{-4}\) instead of \(1 \times 10^{-4}\).

\subsection{Results and Analysis}

\subsubsection{BB Codes}

\begin{figure*}[!htbp]
  \centering
  \begin{subfigure}[t]{0.24\linewidth}
    \centering
    \includegraphics[width=\linewidth]{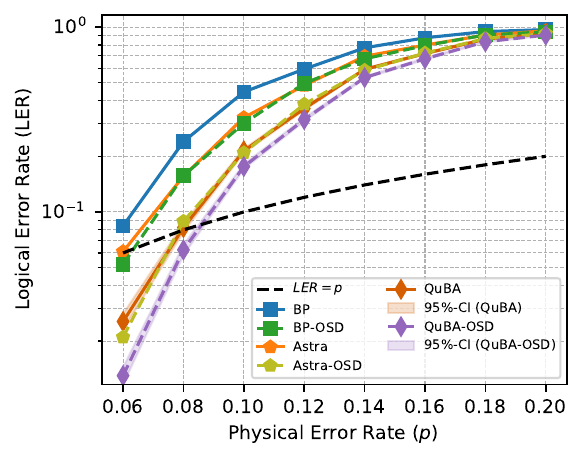}
    \subcaption{$[[90,8,10]]$}
  \end{subfigure}
  \hfill
  \begin{subfigure}[t]{0.24\linewidth}
    \centering
    \includegraphics[width=\linewidth]{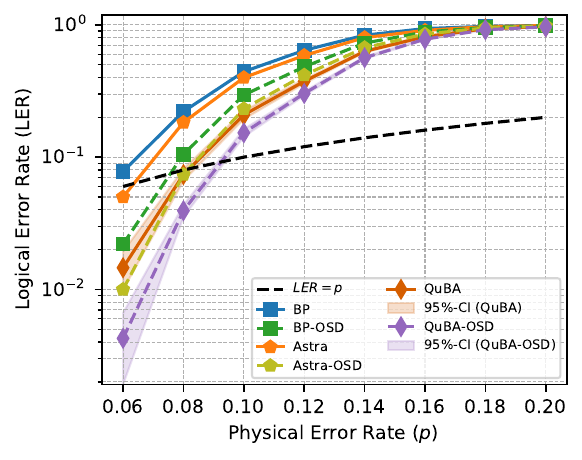}
    \subcaption{$[[144,12,12]]$}\label{bb144}
  \end{subfigure}
  \hfill
  \begin{subfigure}[t]{0.24\linewidth}
    \centering
    \includegraphics[width=\linewidth]{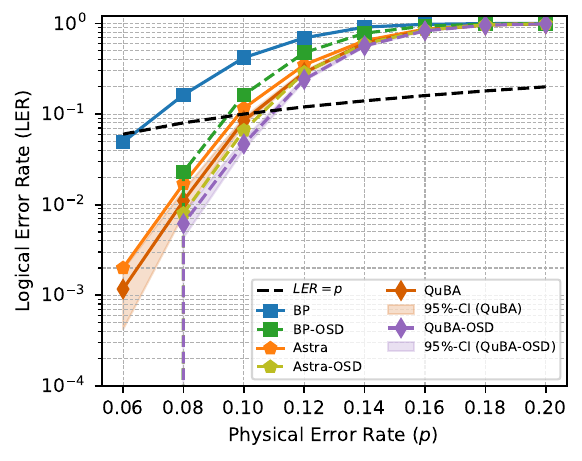}
    \subcaption{$[[288,12,18]]$}
  \end{subfigure}
  \hfill
  \begin{subfigure}[t]{0.24\linewidth}
    \centering
    \includegraphics[width=\linewidth]{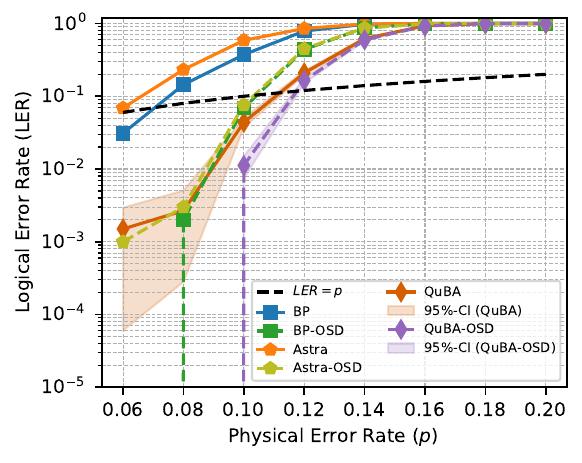}
    \subcaption{$[[756,16,\leq34]]$}
  \end{subfigure}

  \caption{Logical vs. physical error rate (LER vs. $p$) for
  BB codes. Lines become vertical when all errors are corrected (LER=0).}
  \label{fig:bb_all_methods}
\vspace*{-\baselineskip}
\end{figure*}

\textbf{Overall trends -}
Across the BB family, larger codes shift the LER curves downward at
fixed PER and move more QuBA/QuBA-OSD points below the break-even line
(Fig.~\ref{fig:bb_all_methods}), with reductions from the
\(10^{-2}\) regime on \([[144,12,12]]\) to the \(10^{-4}\) regime on
\([[288,12,18]]\) and further gains on
\([[756,16,\leq34]]\). This size-dependent shift indicates that the
decoder exploits larger code distances rather than merely fitting a
fixed-noise regime. For example, \([[90,8,10]]\) remains above
\(10^{-2}\), while \([[144,12,12]]\) already drops below that level,
and larger codes push more PER values below \(LER=p\).

\textbf{Comparison with BP and BP-OSD:}
Across all BB codes, QuBA consistently outperforms BP, with advantages
up to two orders of magnitude. E.g., at \(p=0.06\), QuBA achieves
\(0.00140 \pm 0.00122\) on \([[756,16,\leq 34]]\), compared to BP's
0.031. Under high-confidence decisions (uncertainty lower bound), the
gain increases to two orders of magnitude. Even with OSD
post-processing, QuBA outperforms BP-OSD on smaller codes such as
\([[90,8,10]]\) and \([[144,12,12]]\). For larger codes like
\([[288,12,18]]\) and \([[756,16,\leq 34]]\), QuBA still surpasses
BP-OSD at most PER values, except at \(p=0.06\) where BP-OSD
saturates to zero. For instance, on \([[288,12,18]]\) at \(p=0.10\),
QuBA achieves \(0.08430 \pm 0.00914\), nearly an order lower than
BP-OSD's 0.163. These results highlight that QuBA's robustness
derives primarily from its model architecture rather than reliance on
heavy post-processing.
\textbf{Comparison with Astra:}
Against Astra, QuBA demonstrates similar or even larger
advantages. Across all tested codes, QuBA outperforms Astra, often by
one to two orders of magnitude. E.g., at \(p=0.06\) on
\([[756,16,\leq 34]]\), QuBA achieves \(0.00140 \pm 0.00122\),
compared to Astra's 0.07, a difference of nearly two orders. Under
confidence-based evaluation, this gap widens to nearly 7\%, even
surpassing Astra-OSD (0.001). With OSD, QuBA establishes superiority
over all baselines. For instance, on \([[756,16,\leq 34]]\) at
\(p=0.08\), QuBA-OSD fully converges with no uncertainty
(\(0.00000 \pm 0.00000\)), compared to Astra-OSD's 0.003.

\textbf{Summary:}
Overall, QuBA improves over BP and Astra across all BB codes, with
advantages ranging from one to three orders of magnitude depending on
PER and evaluation setting. With OSD, QuBA-OSD gives the strongest
benchmarks, while QuBA without OSD remains competitive against
post-processing-enhanced baselines, underscoring the strength of its
attention design.

\subsubsection{SAGU - Domain Generalization}

\begin{figure}[!htbp]
    \centering
    \begin{subfigure}[t]{0.49\linewidth}
    \centering
    \includegraphics[width=\linewidth]{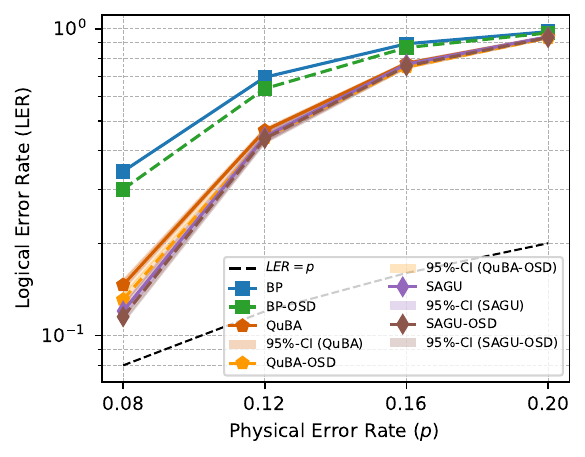}
    \subcaption{$[[72,12,6]]$}
    \end{subfigure}
    \begin{subfigure}[t]{0.49\linewidth}
    \centering
    \includegraphics[width=\linewidth]{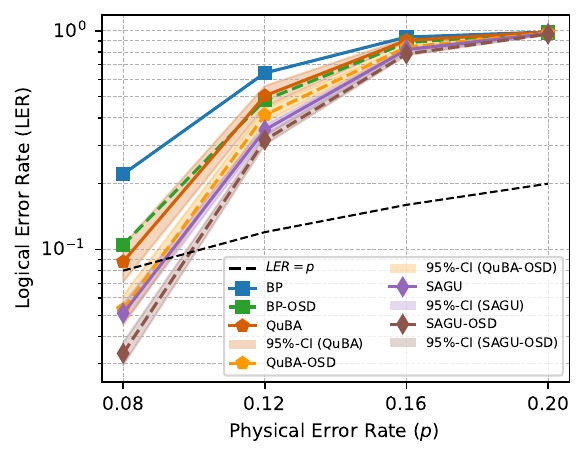}
    \subcaption{$[[144,12,12]]$}
    \end{subfigure}
    
    \begin{subfigure}[t]{0.49\linewidth}
    \centering
    \includegraphics[width=\linewidth]{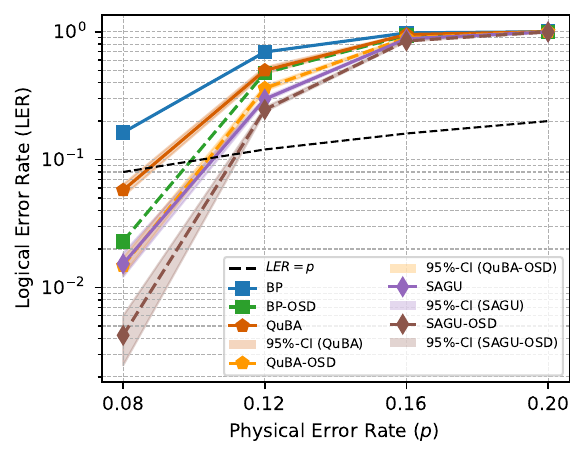}
    \subcaption{$[[288,12,18]]$}
    \end{subfigure}
    \begin{subfigure}[t]{0.49\linewidth}
    \centering
    \includegraphics[width=\linewidth]{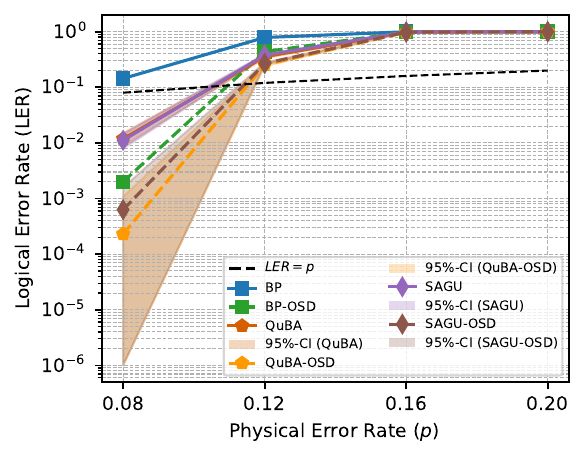}
    \subcaption{$[[756,16,\leq34]]$}
    \end{subfigure}
  
    \caption{Performance comparison of SAGU and other decoding methods
      across all domains, with and without OSD, on BB codes. (a):
      starting domain $[[72, 12, 6]]$; (b): diversity domain
      $[[144, 12, 12]]$; (c): aggregation domain $[[288, 12, 18]]$;
      and (d): out-of-domain $[[756, 16, \leq34]]$.}
    \label{fig:SAGU_bb}
\end{figure}

\textbf{Overall trends:} Across domains, SAGU lowers LER as code size
increases: the starting domain \([[72,12,6]]\) remains above
\(10^{-1}\), the diversity domain \([[144,12,12]]\) falls below
\(10^{-1}\), the aggregation domain \([[288,12,18]]\) reaches about
\(10^{-2}\), and the out-of-domain \([[756,16,\leq34]]\) drops to
\(10^{-6}\) (Fig.~\ref{fig:SAGU_bb}). This nearly four-order
progression preserves the expected rightward break-even shift and
shows that SAGU retains the scaling behavior of domain-specific
training while transferring across domains.

\textbf{Comparison with BP and BP-OSD:}
SAGU consistently outperforms both BP and BP-OSD across the in-domain
codes \([[72,12,6]]\), \([[144,12,12]]\), and \([[288,12,18]]\), with
improvements reaching up to one order of magnitude. For instance, on
\([[288,12,18]]\) at \(p=0.08\), BP attains an LER of 0.163, while
SAGU achieves \(0.01533 \pm 0.00320\). Confidence-based evaluation
further amplifies this advantage, and even under conservative
estimates (upper confidence bounds), SAGU maintains its lead,
underscoring model reliability. With OSD post-processing, SAGU
surpasses all competitors, including BP-OSD and QuBA-OSD, across all
codes. For the larger codes \([[288,12,18]]\) and
\([[756,16,\leq34]]\), at \(p=0.08\), SAGU-OSD achieves an
order-of-magnitude improvement compared to BP-OSD, and outperforms
BP-OSD by two orders of magnitude.
\textbf{Comparison with QuBA:}  
As discussed in previous sections, QuBA already outperforms BP with
and without OSD. Here, we focus on the additional benefits of
cross-domain training with SAGU relative to the domain-specific
QuBA. At \(p=0.08\), SAGU consistently improves upon QuBA by about
0.02–0.03 in LER on \([[72,12,6]]\) and \([[144,12,12]]\), both with
and without OSD, within the confidence bounds. On \([[288,12,18]]\),
SAGU achieves nearly 0.04 improvement over QuBA, and when enhanced
with OSD, SAGU-OSD gains a full order of magnitude advantage over
QuBA-OSD (\(0.00423 \pm 0.00177\) vs.\ \(0.01480 \pm 0.00233\)). On
the out-of-domain code \([[756,16,\leq34]]\), SAGU performs only
marginally worse than QuBA, with differences confined to the fourth
decimal place, while SAGU-OSD and QuBA-OSD show nearly identical
performance.

\textbf{Summary:}
SAGU consistently improves over BP and BP-OSD on in-domain codes and,
with OSD, surpasses QuBA-OSD on nearly all benchmarks. Compared with
domain-specific QuBA, SAGU gives modest but consistent gains on
smaller and intermediate codes while retaining competitive
out-of-domain performance. Overall, these results highlight SAGU's
cross-domain generalization, reliability, and scalability for quantum
LDPC decoding.

\subsubsection{Coprime BB Codes}
\label{app:coprime_bb_code}
\begin{figure}[htbp]
  \centering

  \begin{subfigure}[t]{0.49\linewidth}
    \centering
    \includegraphics[width=\linewidth]{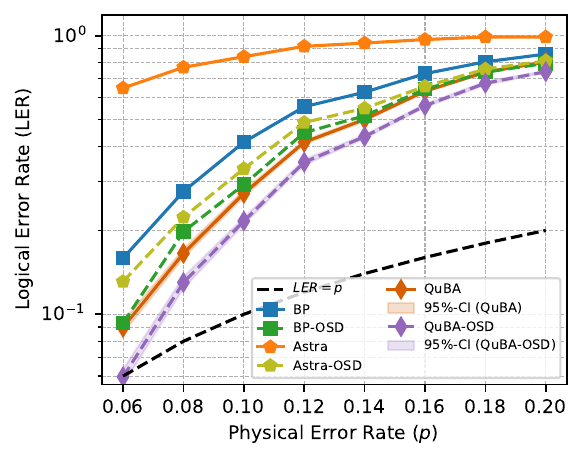}
    \subcaption{$[[30,4,6]]$}
  \end{subfigure}
  \begin{subfigure}[t]{0.49\linewidth}
    \centering
    \includegraphics[width=\linewidth]{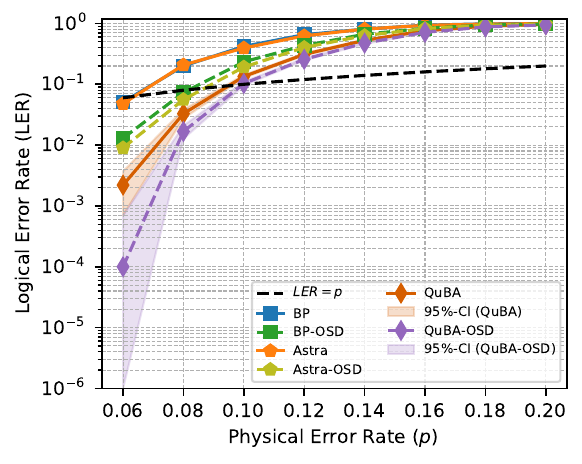}
    \subcaption{$[[154,6,16]]$}\label{copbb154}
  \end{subfigure}

  \caption{Logical vs. physical error rate (LER vs. $p$) for
    coprime BB codes.}
  \label{fig:coprime_all_methods}
\end{figure}

For coprime BB codes, QuBA and QuBA-OSD follow the same size-scaling
behavior observed for standard BB codes: LER drops from the
\(10^{-1}\) regime on \([[30,4,6]]\) to nearly \(10^{-6}\) on
\([[154,6,16]]\) (Fig.~\ref{fig:coprime_all_methods}). Correspondingly,
more PER points fall below the break-even line as code size increases,
matching the behavior of standard BB codes. On
\([[154,6,16]]\) at \(p=0.06\), QuBA maintains approximately an
order-of-magnitude advantage over BP and BP-OSD, and
confidence-bound evaluation widens its margin over Astra to nearly
two orders of magnitude. This advantage also holds against Astra-OSD,
showing that the confidence-bound gains are not only a consequence of
classical post-processing. With OSD, QuBA-OSD exceeds BP-OSD and
Astra-OSD by roughly one order of magnitude and reaches a zero LER
lower bound at \(p=0.06\). These results mirror the standard BB-code
trends and support QuBA's robustness across related code
constructions.

\subsubsection{BB vs.\ Coprime BB Codes}
\label{app:compa_bb_coprime}
It is instructive to compare standard BB codes with their coprime
counterparts at similar distances and comparable block lengths, as this
sheds light on whether or not the algebraic simplifications inherent in
coprime constructions influence
decoder performance.

For BB \([[144,12,12]]\) (see Fig.~\ref{bb144}) and coprime BB
\([[154,6,16]]\) (see Fig.~\ref{copbb154}), decoding results without
OSD are largely comparable across all methods. E.g., at \(p = 0.06\),
BP yields LERs of 0.078 and 0.051, respectively, which are of the same
order of magnitude. For QuBA, the BB code achieves
\(1.47 \times 10^{-2} \pm 4.7 \times 10^{-3}\), whereas the coprime BB
code yields \(2.20 \times 10^{-3} \pm 1.5 \times 10^{-3}\).
\textbf{Effect of OSD:} With OSD post-processing, all decoding methods
again exhibit closely matched performance. For BB \([[144,12,12]]\) at
\(p = 0.06\), the LERs are 0.022 (BP-OSD), 0.01 (Astra-OSD), and
\(4.27 \times 10^{-3} \pm 2.36 \times 10^{-3}\) (QuBA-OSD). For the
coprime BB code \([[154,6,16]]\), the corresponding results are 0.013
(BP-OSD), 0.009 (Astra-OSD), and
\(1.0 \times 10^{-4} \pm 6.0 \times 10^{-4}\) (QuBA-OSD). Under
confidence-bound evaluation, QuBA-OSD converges to zero.

\textbf{Overall:} Both BB and coprime BB codes demonstrate nearly identical
decoding behavior across BP, Astra, and their OSD-enhanced variants,
with differences typically remaining within the same order of
magnitude. Although the coprime BB code attains a lower central value,
the confidence intervals overlap, and both codes fall within a
comparable performance regime. Where separations emerge, they modestly
favor the coprime BB code, consistent with its lower code rate (BB:
0.0833 vs. Coprime BB: 0.0390), which introduces additional stabilizer
constraints and thereby yields a larger code distance (more
correctable errors), as well as structural advantages such as
algebraic simplification.

\subsubsection{Runtime}\label{runtime}

\begin{figure}[htbp]
  \centering

  \begin{subfigure}[t]{0.49\linewidth}
    \centering
    \includegraphics[width=\linewidth]{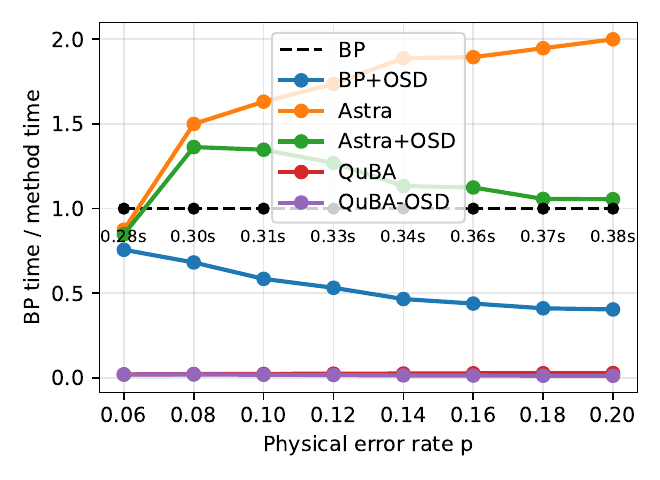}
    \subcaption{$[[90, 8, 10]]$}\label{8:a}
  \end{subfigure}
  \begin{subfigure}[t]{0.49\linewidth}
    \centering
    \includegraphics[width=\linewidth]{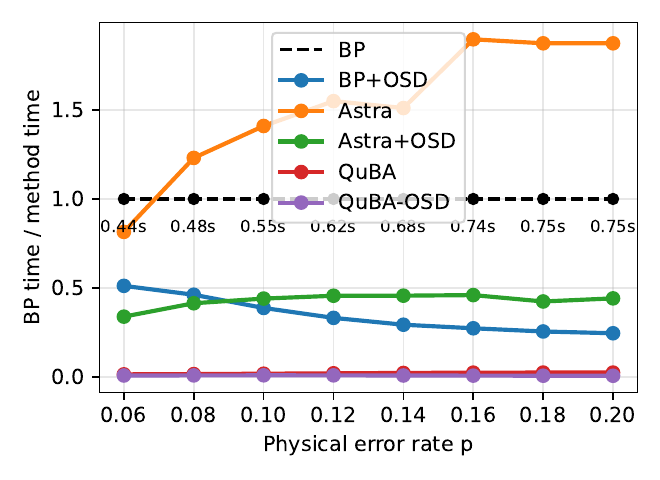}
    \subcaption{$[[154,6,16]]$}\label{8:b}
  \end{subfigure}
    \begin{subfigure}[t]{0.49\linewidth}
    \centering
    \includegraphics[width=\linewidth]{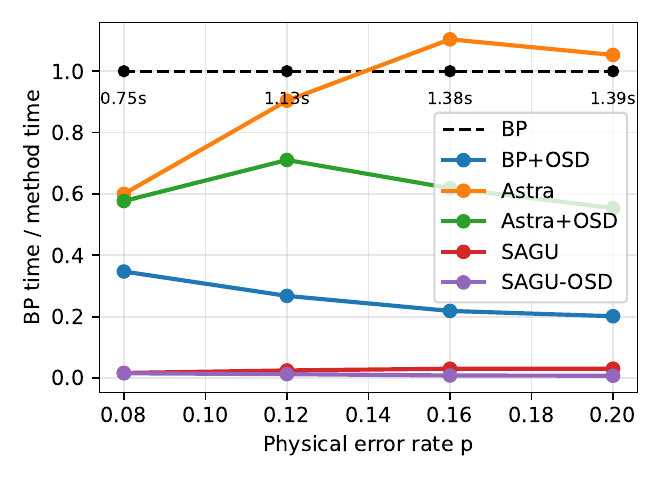}
    \subcaption{$[[288,12,18]]$}\label{8:c}
  \end{subfigure}
  \begin{subfigure}[t]{0.49\linewidth}
    \centering
    \includegraphics[width=\linewidth]{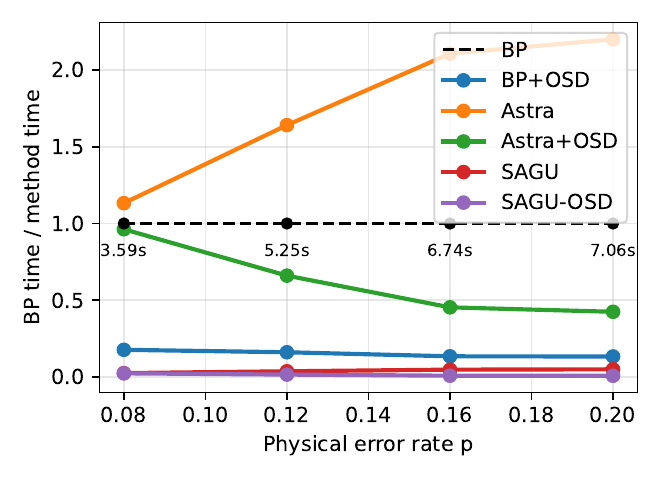}
    \subcaption{$[[756,16,34]]$}\label{8:d}
  \end{subfigure}

  \caption{Runtime relative to the BP baseline across multiple codes
    (one BB code, one coprime BB code, and two BB codes for SAGU
    in-domain and out-of-domain) as a function of the physical error
    rate $p$. Values above (below) 1 indicate faster (slower)
    performance than BP. The dashed horizontal line marks the BP
    baseline, and the labels along that line show the absolute BP
    runtime at each $p$.}
  \label{fig:speedup}
\end{figure}

Fig.~\ref{fig:speedup} reports runtime normalized to BP. Average
runtime increases with larger block sizes within the same code
family. Astra(-OSD) is generally fastest on \([[90,8,10]]\); for
coprime BB codes, BP-OSD is faster than Astra-OSD for
\(p\in[0.06,0.09]\) but slower afterward; under SAGU, QuBA(-OSD)
approaches BP-OSD runtime on larger domains while remaining slower
than Astra(-OSD).

The overhead reflects QuBA's Bayesian inference: Each prediction uses
\(M=30\) MC forward passes, each with \(n_{\text{iters}}\)
message-passing iterations and \(L=7\) Bayesian linear modules, so
stochastic parameter draws scale as
$M \times n_{\text{iters}} \times L$ (or
$M \times n_{\text{iters}} \times 2L$ if biases are counted). This
makes inference approximately \(M\) times more expensive than
deterministic inference, before accounting for Bayesian parameter
sampling inside each pass. The resulting accuracy--runtime tradeoff
improves confidence estimation and decoding quality, but makes the
current implementation less suitable for real-time decoding without
lighter Bayesian inference or hardware acceleration.
Thus, runtime should be interpreted alongside LER gains.

\section{Conclusions}
We presented QuBA, a Bayesian graph neural network decoder that
combines edge-aware attention with recurrent memory, enabling both
uncertainty-aware predictions and effective multi-round
reasoning. Building on this architecture, we introduced SAGU, a
sequential training paradigm that promotes generalization across
quantum LDPC codes and noise regimes. Our experiments on BB and
coprime BB codes demonstrate that QuBA(-OSD) consistently outperforms
classical decoders (BP, BP-OSD) and state-of-the-art neural approaches
(Astra), with gains reaching up to \emph{two orders of magnitude}
accuracy improvement even without considering the confidence-decision
bounds, e.g., for the coprime BB $[[154, 6, 16]]$ code. Notably, these
advantages hold even in the absence of OSD post-processing,
highlighting QuBA’s robustness. Moreover, SAGU generalizes
successfully across domains, maintaining high performance on codes
previously unseen during training. These results highlight the promise
of Bayesian-attention graph networks for scalable quantum
decoding. However, given the runtime limitations discussed in
Sec.~\ref{runtime}, future work will focus on developing a
lightweight, uncertainty-aware decoder under the circuit-level error
model, potentially leveraging convolutional neural
networks~\cite{gu2026scalable} to predict logical failures instead of
physical errors.

\section*{Acknowledgment}
This work was supported in part by NSF awards OSI-2531350,
OSI-2410675, PHY-2325080, and OMA-2120757.

\bibliographystyle{IEEEtran}
\bibliography{paper}

\end{document}